\documentclass[sigconf,authorversion]{acmart}

\AtBeginDocument{%
  \providecommand\BibTeX{{%
    \normalfont B\kern-0.5em{\scshape i\kern-0.25em b}\kern-0.8em\TeX}}}

\usepackage{algorithmic}
\usepackage{graphicx}
\usepackage{xcolor}
\usepackage{algorithm}
\usepackage{multirow}
\usepackage{makecell}
\usepackage{stackengine}
\usepackage{float}
\usepackage{tabularx}
\usepackage{adjustbox}
\usepackage{stfloats}
\usepackage{url}
\usepackage{stackengine}
\usepackage{setspace}
\usepackage{colortbl}
\usepackage{textcomp}
\usepackage{booktabs}
\usepackage[flushmargin]{footmisc}
\usepackage{bbding}
\usepackage{soul}
\usepackage{siunitx}
\usepackage{balance}
\usepackage{flushend}

\begin{document}

\setcopyright{none}
\settopmatter{printacmref=false}
\renewcommand{\footnotetextcopyrightpermission}[1]{}
\pagestyle{empty}

\title{Reuse and Blend:  
Energy-Efficient Optical Neural Network Enabled by Weight Sharing}

\author{Bo Xu}
\authornote{Equal Contribution.}

\affiliation{MICS Thrust, \\The Hong Kong University of Science \\and Technology (Guangzhou)\country{}}
\email{bxu721@connect.hkust-gz.edu.cn}

\author{Yuetong Fang}
\authornotemark[1]
\affiliation{MICS Thrust, \\The Hong Kong University of Science \\and Technology (Guangzhou) \country{ }}
\email{yfang870@connect.hkust-gz.edu.cn}

\author{Shaoliang Yu}
\authornote{Corresponding Author.}
\affiliation{Zhejiang Laboratory \country{  } }
\email{yusl@zhejianglab.com}

\author{Renjing Xu}
\authornotemark[2]
\affiliation{MICS Thrust, \\The Hong Kong University of Science \\and Technology (Guangzhou)  \country{   } }
\email{renjingxu@hkust-gz.edu.cn}



    

\begin{abstract}
Optical neural networks (ONN) based on micro-ring resonators (MRR) have emerged as a promising alternative to significantly accelerating the massive matrix-vector multiplication (MVM) operations in artificial intelligence (AI) applications. However, the limited scale of MRR arrays presents a challenge for AI acceleration. The disparity between the small MRR arrays and the large weight matrices in AI necessitates extensive MRR writings, including reprogramming and calibration, resulting in considerable latency and energy overheads. 
To address this problem, we propose a novel design methodology to lessen the need for frequent weight reloading. Specifically, we propose a reuse and blend (R\&B) architecture to support efficient layer-wise and block-wise weight sharing, which allows weights to be reused several times between layers/~blocks. Experimental results demonstrate the R\&B system can maintain comparable accuracy with 69\% energy savings and 57\% latency improvement. These results highlight the promise of the R\&B to enable the efficient deployment of advanced deep learning models on photonic accelerators.
\end{abstract}

\maketitle

\section{Introduction}
The rapid growth of AI has led to a growing need for efficient computing architecture~\cite{shen2017deep}. However, with the end of Moore's law, the power consumption of integrated circuits has reached a bottleneck in meeting the requirements of state-of-the-art AI algorithms, particularly as the required number of arithmetic operations grows from millions to trillions. As a promising alternative to traditional electronic computing platforms, optical neural networks (ONNs) have recently emerged, offering high bandwidth, low power usage, and massively parallel processing ~\cite{shen2017deep, zhou2023prospects,xu2024large}. Among ONN architectures, micro-ring resonator-based ONNs (MRR ONNs) stand out for their compact design and versatility in both modulating inputs and filtering signals, making them an efficient accelerator for the massive matrix-vector multiplication (MVM) in AI algorithms. This paves the way for ultra-dense computing with massive wavelength division multiplexing (WDM) links~\cite{
feldmann2021parallel}.

However, the limited size of a single MRR array compared to the large weight matrices in neural networks necessitates frequent weight reprogramming for MRR ONN-based AI acceleration, subsequently raising the degradation issues of MRR devices. The most common MRR programming techniques depend on thermal-optical modulators and intricate calibration processes, which result in significant reconfiguration overheads, increased delays, and higher energy consumption. MRR writing involves calibrating a weight-current response curve for each wavelength channel and programming the desired value based on initial search points. However, temperature fluctuations can easily distort these response curves, making MRR writing processes sensitive and unstable. While integrating proportional-integral-differential (PID) control loops can help mitigate this issue through additional reprogramming efforts, they are time-consuming and require extra high-speed peripheral circuits, which have become a bottleneck in the system throughput. Besides, the energy consumed by the control mechanism increases with modulation frequency, making high-speed MRR ONNs less energy-efficient.

\begin{figure}[t]
    \centering
    \includegraphics[width=1\linewidth]{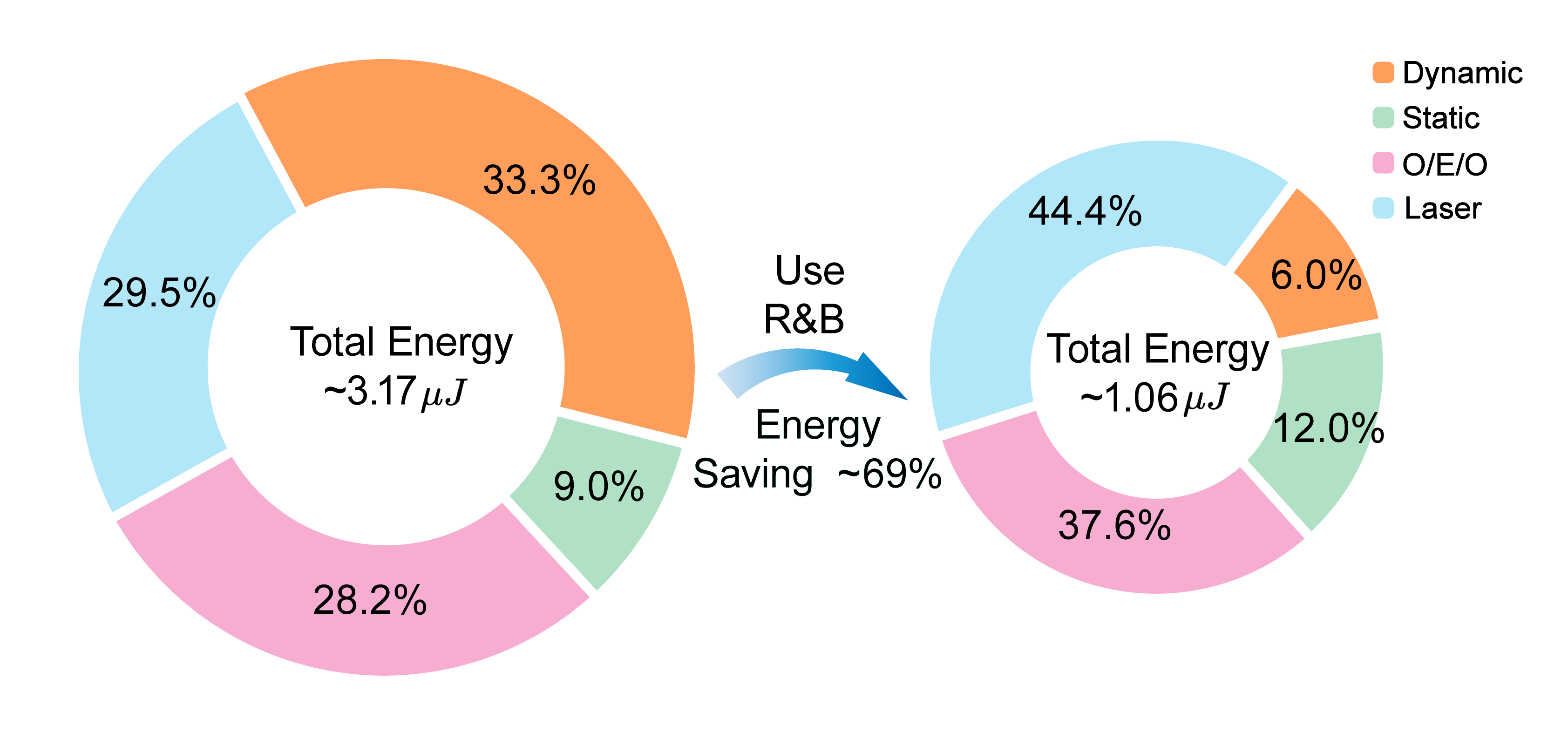}
    \caption{
Comparison of energy consumption: without weight sharing (left) vs. with weight sharing (right). The reuse-and-blend (R\&B) method significantly reduces dynamic power consumption, mainly due to decreased MRR writing frequency and, thus, lower energy needed for programming and calibration.
    }
    \label{fig1}
\end{figure}

\begin{figure*}
    \centering
    \includegraphics[width=1\linewidth]{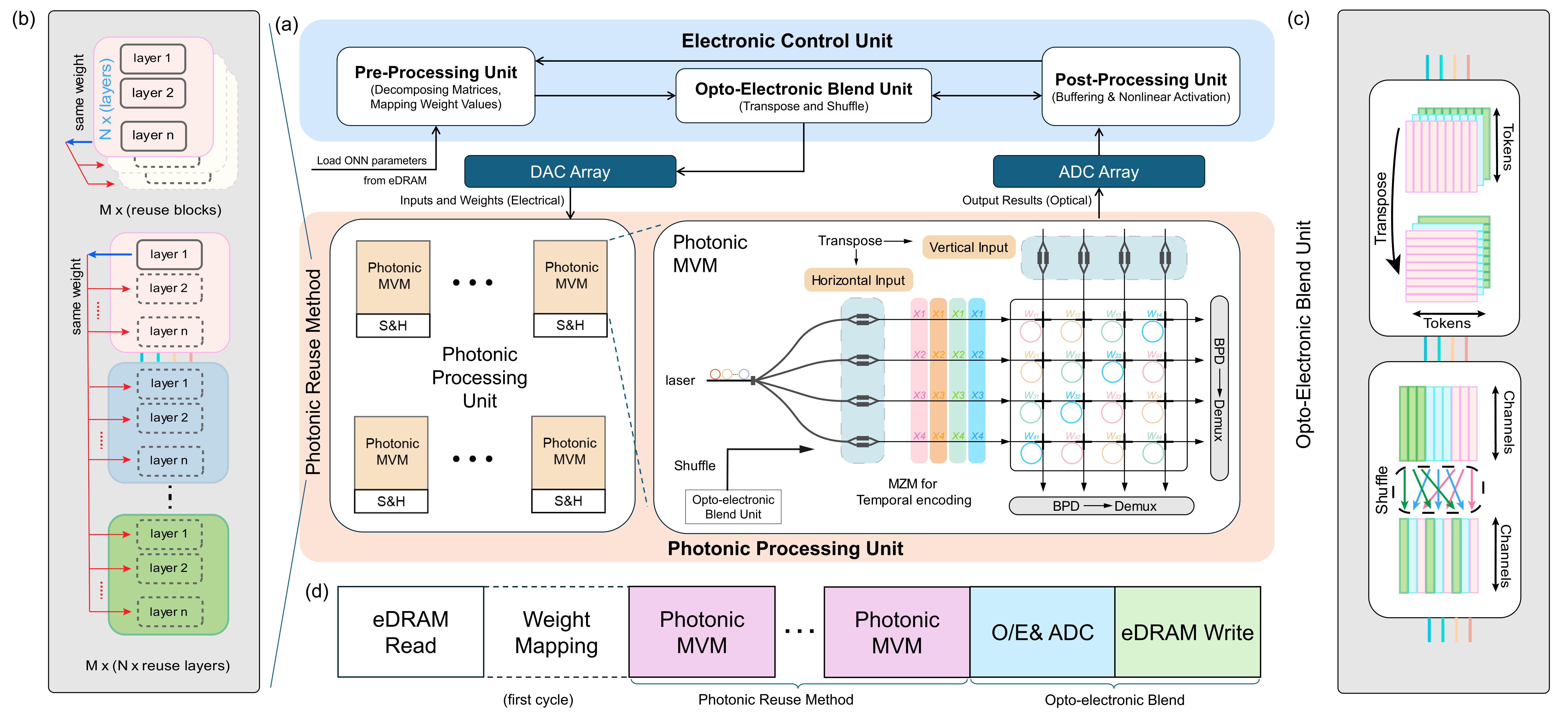}
    \caption{ (a) Overview of the R\&B architecture. Each PPU contains a photonic MVM unit and a sampling and hold (S\&H) unit. (b) Photonic Reuse Method (PRM). Block-wise reuse allows weight sharing among blocks (a block typically contains multiple layers). Layer-wise reuse enables weight sharing between individual layers. (c) Opto-electronic Blend Unit (OBU). OBUs handle shuffle operations via the peripheral circuit and perform transpose operations directly in the optical domain. (d) Computing pipeline of our R\&B architecture. }
\label{main}
\end{figure*}

Although MRR arrays consume minimal power during MVM operations, frequent MRR writings lead to significant dynamic power consumption and delays, especially when high computing bandwidth is needed (Fig.\ref{fig1}). Prior works have focused on optimizing MRR ONN energy efficiency and throughput by enhancing data locality~\cite{RN30}, reducing computations through optimized MVM algorithms~\cite{mehrabian2019winograd}, or adopting advanced MRR devices with algorithm-hardware co-design~\cite{gu2022squeezelight}. However, the limited scale of MRR arrays makes intra- and inter-layer weight remapping inevitable in current MRR ONN designs. {For example, a $8\times8$ MRR array is deemed large-scale due to crosstalk limitations~\cite{RN11}, while the largest convolutional layer in ResNet-18 can have a weight matrix of dimensions $512\times512\times3\times3$~\cite{he2016deep}. This limitation hinders the implementation of state-of-the-art networks on MRR-based platforms.

To address these challenges, we propose an efficient {R\&B} MRR ONN architecture that enables layer/block-wise weight sharing. Our novel Photonic Reuse Method (PRM) prolongs the life cycle of each weight value, significantly reducing programming and calibration operations during MRR ONN inference. Besides, our Opto-electronic Blend Unit (OBU) enhances feature extraction through feature transpose and channel shuffle. By adopting our weight-sharing technique, we achieved a 69\% reduction in energy consumption and a 57\% improvement in latency. The main contributions are as follows:

\begin{itemize}
\item We propose a weight-sharing technique that significantly reduces the programming and calibration needed across blocks/layers. This provides a generalizable method to save dynamic energy consumption that can be applied not just to optical computing but also to other novel computing paradigms. 
\item The R\&B MRR-ONN architecture significantly improves performance with negligible overhead. It seamlessly combines optical and electronic processing with an optoelectronic blend unit to enable efficient optical transposing and electronic shuffling.
\item   Extensive experiments validate that our weight-sharing techniques R\&B architecture can effectively reduce the need for reprogramming and calibration processes. By implementing our weight-sharing network, we achieved a 69\% reduction in energy consumption and a 57\% improvement in latency.
\end{itemize}


\section{Background}

\textbf{Microring Resonators} (MRRs) use ring-shaped waveguides to selectively couple lights at specific wavelengths between the ring and nearby bus waveguides. MRR crossbars accelerate matrix-vector multiplications by weighting signals at specific resonance wavelengths, and summing WDM links, whereby encoding the input vectors and forming partial dot products. This makes MRR ONN a promising technology for AI acceleration~\cite{xu202111, tait2017neuromorphic}. \textbf{However, long-term operation of MRRs under constant or varying temperatures and voltages can cause significant aging problems.} This becomes critical for real-world use since the most common MRR programming relies on heat modulation\cite{Zheng:14}. Furthermore, the thermal sensitivity of mainstream programming techniques makes MRR writings unstable, which takes up a large proportion of extra dynamic energy for reprogramming and calibration. To address this issue, our work investigates the PRM that greatly reduces the required frequency of MRR writing. We also propose an R\&B MRR-ONN architecture with the OBUs. These two techniques jointly improve energy efficiency, system throughput and performance with negligible overhead.

\vspace{5.5mm} 
\noindent\textbf{B$\&$W protocol} has been widely employed for implementing MRR-based ONNs based on the WDM technology, using a bank of tunable MRRs to represent synaptic weights and broadcasting an input vector to all MRRs to perform on-chip MVM operations. DEAP in ~\cite{RN10} has proposed a convolutional neural network (CNN) hardware architecture that combines digital electronics and analog photonics, utilizing voltage addition for accumulating partial dot products across filter channels and achieving up to 14 times acceleration compared to GPUs. Cheng et al. proposed the hitless weight-and-aggregation architecture in ~\cite{RN28}, which helps to reduce thermal crosstalk and improve weight resolution. In ~\cite{RN29}, Bagherian et al. reduced the patching cost in CNNs by adopting optical delay lines. Later in ~\cite{RN30}, Xu et al. proposed to improve the scheme by reusing one single optical delay line, thus reducing the required modulation resources.  

\textbf{ However, the SOTA MRR-ONNs have struggled to accelerate large MVMs, which are necessary for advanced networks.} The limited scale of a single MRR array means that weight remapping within and between layers is still necessary in current MRR-ONN designs, which has prevented these designs from being used for advanced network deployment. Our work proposes an efficient weight-sharing approach that greatly reduces the need for frequent reprogramming and calibration. This enables MRR-based AI acceleration to reach its full potential.

\section{R\&B Architecture}

The R\&B architecture, depicted in Fig.~\ref{main}, includes multiple Photonic Processing Units (PPUs), each containing a photonic MVM unit paired with a sample-and-hold (S\&H) unit for data retention. A key innovation is the PRM, which schedules MRR writes and efficiently shares weights across blocks or layers, reducing reprogramming and calibration times. This enables a single MRR array to represent multiple weight matrices, thereby extending the lifespan of each weight value. Supported by OBUs, the weights from one MRR array can be dynamically transformed through optical transpose and electrical shuffle.  These techniques enable the R\&B architecture to significantly lower the MRR writing frequency, enhancing operational efficiency and reducing power consumption.



\begin{algorithm}
    \renewcommand{\algorithmicrequire}{\textbf{Input:}}
    \renewcommand{\algorithmicensure}{\textbf{Output:}}
    \caption{R\&B architecture}
    \label{alg1}
    \begin{algorithmic}[1]
        \REQUIRE {$x \in \mathbb{R}^{c \times h \times w}$, T}
        \STATE A R\&B network :
        $\mathcal{N}_M: \left[b_1, b_2 \ldots, b_m, \ldots, b_M\right]$
        \FOR {$b_{m}$ in ${N}_M$}
            \IF {$b_{\text {reuse }}$}
                \STATE reuse\_times = T
                \FOR {$i$ in $1,2,..., reuse\_times$}
                \IF {Shuffle} 
                    \STATE $b_{\text {reuse }} \in \mathbb{R}^{c \times h \times w} \rightarrow b_{\text {reuse }}^{\prime} \in \mathbb{R}^{\frac{c}{g} \times g \times h \times w} \rightarrow b_{\text {reuse }}^{\prime \prime} \in \mathbb{R}^{c \times h \times w}$
                \ELSIF{Transpose}
                    \STATE $b_{\text {reuse }} \in \mathbb{R}^{c \times h \times w} \rightarrow b_{\text {reuse }} \in \mathbb{R}^{c \times w \times h}$
                
                \ENDIF            
                \STATE $x = b_{\text {reuse }} \cdot x$
                \ENDFOR 
            \ENDIF
        \ENDFOR
    \end{algorithmic}  
\end{algorithm}

\vspace{{-4mm}}
\subsection{Photonic Reuse Method}

As shown in Fig.~\ref{main}(a), the PRM is supported by the PPUs, each containing an MRR cross-bar array~\cite{Zheng:14}. Input data is encoded into 1D sequential vectors and converted to optical signals by lasers. These signals are then processed by PPUs, which function as tensor cores. The generated partial dot products are converted back to the electrical domain by balanced photodetector (BPD) arrays. Finally, the results are processed by trans-impedance amplifiers, low-pass filters, and Analog-to-Digital Converters (ADCs) to produce the final output.

In neural network architectures, particularly those involving intensive computations, the frequent updating of weights during forward propagation can result in significant dynamic energy consumption. To address this challenge, implementing weight-sharing strategies across different blocks or layers could be effective, as illustrated in Fig.~\ref{main}(b). This approach minimizes the need to continuously update weights, thereby conserving energy. We propose the PRM for R\&B architecture, which supports two types of reuse granularity: Layer-wise Reuse and Block-wise Reuse.

An M-block R\&B network $\mathcal{N}_M$ contains a series of blocks:


\vspace{-0.2cm}
\begin{equation}
\mathcal{N}_M: \left[b_1, b_2, \ldots, b_m, \ldots, b_M\right] \\
\end{equation}
\noindent where $b_m$ represents the m-th block within the R\&B architecture, which consists of $N$ layers:

\vspace{-0.4cm}
\begin{equation}
b_m: l=\left[l_1, l_2, \ldots,l_n, \ldots, l_N\right] 
\end{equation}

PRM sets a basic weight block (such as the mixer block in MLP-Mixer or the residual block in ResNet) as the reuse weight matrix for each PPU, with $B$ signifying the basic weight block space:
\begin{equation}
\begin{split}
B_{\text {reuse }} &= [b_{\text {reuse-1 }}, b_{\text {reuse-2 }}, \ldots, b_{\text {reuse-R }}]; \quad B_{\text {reuse }} \in \mathrm{B} \\
\end{split}
\end{equation}


In block-wise reuse, the reused block utilizes the weight values from the same basic weight block, along with dynamic transformations enabled by OBUs (Sec. 3.2). For instance, if blocks $b_{\text {m}}$ to $b_{\text {m+p}}$ are scheduled through block-wise weight-sharing strategy, with their basic weight block denoted as $b_{\text {reuse-m}}$ then:

\vspace{-0.2cm}
\begin{equation}
\begin{aligned}
    &\left[b_m, \ldots, b_{m+p}, \ldots,  b_{m+P} \right]= \\ &\left[b_{\text {reuse-m }}^1, \ldots,  b_{\text {reuse-m }}^p, \ldots,  b_{\text {reuse-m }}^P \right]
\end{aligned}
\end{equation}


\noindent The block $b_{reuse-m}$ is reused across $P$ weight-sharing blocks, and $b_{reuse-m}^{p}$ is the $p$-th transformed basic weight layer generated by OBU. Similarly, when a block contains only one layer, block-wise reuse can be considered layer-wise reuse:

\vspace{-0.2cm}
\begin{equation}
\left[l_n, \ldots, l_{n+q}, \ldots,  l_{n+Q} \right]= \left[l_{\text {reuse }}^1, \ldots,  l_{\text {reuse }}^q, \ldots,  l_{\text {reuse }}^Q \right]
\end{equation}

Algorithm 1 presents the PRM workflow. PRM enables reuse of the weight values assigned to each PPU as much as possible thereby extending the lifespan of each weight value. This helps to reduce the model size and the need for reprogramming while maintaining competitive accuracy. Our PRM method is versatile and not limited to optical computing.

\begin{figure}[t]
    \centering
    \includegraphics[width=1\linewidth]{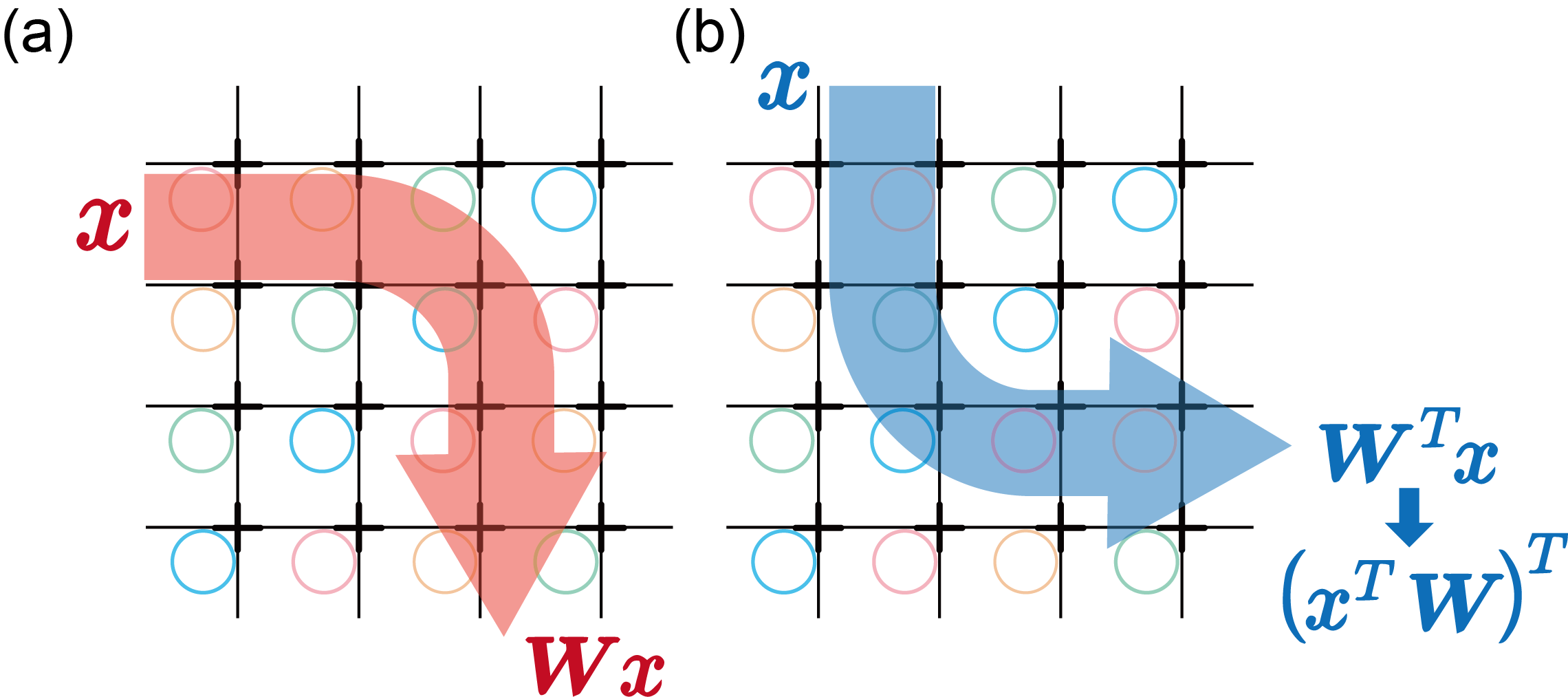}
    \caption{ Illustration of transposed dot production in MRR: (a) horizontal input and (b) vertical input.}
\label{fig3}
\end{figure}

\subsection{Opto-electronic Blend Unit}

The OBU allows each MRR array in R\&B architecture to represent multiple weight matrices with negligible overhead. The OBU consists of two main components: (1) Direct transpose and (2) Random shuffle of the intermediate outputs across layers and blocks, as illustrated in Fig.~\ref{main}(c). This helps maintain accuracy while significantly saving the dynamic energy consumption and latency with the PRM method. By blending channels or tokens, the original basic weight block/layer can be subjected to diversified weight matrices through nonlinear transformations. This enhance the expressiveness and efficiency of the model.

Within the R\&B architecture, specifically considering a layer with output dimensions of C $\times$ H $\times$ W, two innovative shuffling methods are employed: The first generates a random index for the flattened output grouped into blocks, then reorders it to achieve shuffling; The second method divides the output channels into groups and executes shuffling within each group. These rearrangement methods enhance the network's ability to extract a more varied and complex array of features from the input data by mixing features across different channels or tokens. We utilize the necessary optical-to-electrical (O/E) conversion process in MRR ONN to achieve the shuffle operation. The use of BPDs for O/E conversion can ensure high-bandwidth computations, capable of achieving speeds exceeding 30 Gb/s~\cite{RN35}

As illustrated in Fig.~\ref{main}(a), OBUs can perform the transpose operation entirely within the optical domain. This capability allows the weights, denoted as $ \boldsymbol{W_b}$ and $ \boldsymbol{W}_b^{\text{T}}$, to align seamlessly with the horizontal and vertical laser inputs. By enabling transpose operations, OBUs facilitate the mixing of features across various spatial locations, also enhancing feature extraction between channels. Fig.~\ref{fig3} further illustrates the optical signal path involved in this operation. Adding MZMs and BPDs in our R\&B network at the inputs and outputs not only provides significant energy savings but also reduces latency, all with negligible area overhead. This integration underscores the efficiency and performance benefits of using OBUs in advanced optical networks.

The combination of OBU and PRM enhances both accuracy and energy efficiency, allowing the dynamic generation of multiple different weight matrices from a single fundamental matrix and reducing the programming frequency of MRRs. This process significantly enriches the performance and robustness of the model, facilitating advanced and efficient processing capabilities within the R\&B architecture.

\begin{table}[h]
\renewcommand{\arraystretch}{0.6}
\centering
\caption{Main Properties of Component in Our Architecture}
\small
\resizebox{0.45\textwidth}{!}{%
    \begin{tabular}{ccc}
    \cmidrule{1-3}
    \multicolumn{2}{c}{ Components } &  { Value } \\ \cmidrule{1-3}
    \multirow{4}{*}{Waveguide} & $w \times h$  &$\begin{array}{c} 400 \times 300 \mathrm{~nm} \\  \left(h_{\text {slab }}=100 \mathrm{~nm}\right)\end{array}$\\ \cmidrule{2-3}
    & $n_{e f f}, n_g$ & $\begin{array}{c}2.845,3.975 \\
    \text { @ } \lambda=1310 \mathrm{~nm}\end{array}$ \\ \cmidrule{1-3} 
    \multirow{10}{*}{ MRR } & Radius & $6 \mu \mathrm{m}$ \\ \cmidrule{2-3}
    & $\kappa^2$ & 0.04 \\ \cmidrule{2-3}
    & FSR & $11.3 \mathrm{~nm}$ \\ \cmidrule{2-3}
    & Cell area & $127 \times 127 \mu m^2$ \\ \cmidrule{2-3}
    & Modulator Driver &0.8$mW\text { @ }10Gbps$ \\ \cmidrule{2-3}
    & Heater and Tuner power & 14$mW$ \\ \cmidrule{1-3}
    TWMZM  & Bandwidth & $36 \mathrm{GHz} @ 1 \mathrm{~V}$ \\ \cmidrule{1-3}
     Laser & Linewidth & $5 \mathrm{MHz}$ \\
    \cmidrule{1-3}
    \multirow{2}{*}{ PD } & Responsivity & $1.1 \mathrm{~A} / \mathrm{W}$ \\ \cmidrule{2-3}
    & Dark current & $25 \mathrm{pA} \text { @ } 1 \mathrm{~V}$ \\ \cmidrule{1-3}
    ADC &Area & $1.2288 mm^2\text{@}39mW$ \\ \cmidrule{1-3}
    DAC &Area & $0.0004 mm^2\text{@}3.93mW$ \\ \cmidrule{1-3}
    S$\&$H &Area & $0.00004 mm^2$ \\ \cmidrule{1-3}
    eDRAM &Area & $0.268 mm^2$ \\ \cmidrule{1-3}
    Bus &Area & $0.009 mm^2$ \\ \cmidrule{1-3}
    \end{tabular}
}

\label{table1}
\end{table}

\vspace{-5pt}
\subsection{Computing Pipeline}  
The computation process of our R\&B architecture contains three stages, as illustrated in Fig.~\ref{main}(a). Initially, inputs are retrieved from eDRAMs, and corresponding weights are allocated to MRRs. Subsequently, following the PRM configuration, the weights are fixed and reused, allowing the inputs to pass through the MRRs to be optically weighted. The intermediate MVM results generated by the PPUs are then detected by BPDs, where they are converted into summed currents and digitized by ADCs. In the final stage, OBUs transform these outputs to generate the layer-wise results, which are then stored back in eDRAMs in preparation for the next computational layer.



A critical aspect of this architecture is the role of the OBU during inference, mirroring its function during training and inference by executing essential shuffle and transpose operations. Subsequently, following the PRM configuration, the weights are fixed and reused. Along with PRM, these two technologies constitute the primary innovation of our R\&B architecture. By leveraging one MRR array to represent multiple weight matrices, the architecture dramatically reduces the frequency of MRR writing operations, along with power consumption and latency, all while sustaining high performance. 

\vspace{{-2mm}}
\subsection{Hardware Feasibility}
\subsubsection{MRRs Array  for Full-range Weights}
To enable advanced networking capabilities, it was necessary to facilitate matrix-vector multiplication (MVM) incorporating negative values. To address this, we use a strategy involving an offset matrix, denoted as $\small \boldsymbol{W}_{\text {o }}$ \cite{ohno2022si}. The process involves modifying $\small \boldsymbol{W}_{\text {b }}^{\prime}$ as follows:
\begin{equation}
\boldsymbol{W}_{\text {b }}^{\prime}=\frac{1}{2} \boldsymbol{W}_{\text {b }}+\boldsymbol{W}_{\text {o }}
\end{equation}
$\small\boldsymbol{W}_{\text {o }}$ is defined as a matrix with uniformly positive elements. After normalization, the values in $\small\boldsymbol{W}_{\text {b }}$ are confined to the range [-1, 1], and the elements in the transformed matrix $\small\boldsymbol{W}_{\text {b }}^{\prime}$ fall within the [0, 1] range. Subsequently, we perform MVM using both $\small\boldsymbol{W}_{\text {o}}$ and $\small\boldsymbol{W}_{\text {b}}^{\prime}$ against an input vector $x$. This results in the outputs $\small \boldsymbol{W}_{\text {o }} x$ and $\small\boldsymbol{W}_{\text {b }}^{\prime} x$, respectively. Given the uniform nature of $\small \boldsymbol{W}_{\text {o}}$, the size of the MRR crossbar array required for its calculation can be minimized. The final expression for $\boldsymbol{W}_{\text {b}}$ is obtained as follows:$\small W_{\text {b }} x=2(W_{\text {b}}^{\prime} x-W_{\text {o}} x)$. 

The hardware resource introduced by $\boldsymbol{W}_{\text {o}}$ is negligible, particularly for large $N$, with only 1 × $N$ MRR devices are required for computing the MVM of $\boldsymbol{W}_{\text {o}}$. To align with the all-positive signaling scheme in optical systems, we use the ReLU activation function to eliminate negative values in each layer.


\subsubsection{Nonlinear Activation} The ReLU nonlinear activation function is applied in the electrical domain, transforming the output of all neurons to be positive.
\subsubsection{Normalization} Normalization layers, such as BatchNorm, are efficiently implemented using the TIA gain and voltage signal offsets without introducing any additional latency.

\subsubsection{Shuffle and Transpose Operation in OBU}
OBUs handle shuffle operations in the electrical domain and require an additional set of MZMs and BPDs for input modulation and output detection, enabling direct optical transpose operations with negligible area overhead. This is supported by existing techniques, with BPDs achieving speeds over 30 Gb/s~\cite{RN35} and MZMs exceeding 100 Gb/s~\cite{Wang:21}.

\section{Experimental Result}
We evaluate our methods on several modern architectures, including MLP, VGG-13, Resnet-18, and MLP-Mixer on MNIST, CIFAR10 and CIFAR100. We conduct all the experiments in PyTorch and evaluate the performance on a machine with an AMD EPYC 7542 cpu and an NVIDIA RTX 4090 GPU. In the reuse-and-blend training flow, we use an Adam optimizer with an initial learning rate (lr) of 0.001 and a weight decay of 0.01. We use a cosine-annealing lr scheduler during training. All models are trained for 200 epochs except MLP, which is trained for 100 epochs. The models are then quantized to 8-bit for both weights and activations as general setting ~\cite{li2021brecq}.

We consider an MRR array of $8 \times 8$ scale as the basic photonic MVM tile in the following evaluation. 
Table~\ref{table1} summarizes the main characteristics of the photonics components in our evaluation.

\subsection{System Performance Evaluation}
\subsubsection{Compute efficiency and Delay Analysis} Most MRR ONNs require additional nonlinear mapping to encode inputs/weights into voltage signals:

\begin{equation}
\centering
v_x = \sqrt{\phi^{-1}\left(f^{-1}(x)\right)}
\end{equation}

The non-linear mapping process greatly complicates programming.  It often requires multiple write attempts to approach the target programming value, thereby incurring substantial delays and energy consumption, which accounts for approximately 33.3\% of the total energy expenditure. Table~\ref{table2} summarizes the quantitative results of computation efficiency and delay analysis. We use  K matrices with dimensions of $M \times N$, C determines the loop duration for mapping and calibrating a weight, and B is the DWDM capacity. The area ratio $\beta_a$ and power ratio $\beta_p$ between one MZI and one MRR are $\beta_a$=24 and $\beta_p$=12. For CrossLight~\cite{9586161}, the thermal eigen decomposition is expressed as a ratio of $\beta_t$.  

In an advanced network, scaling up both K and C inevitably leads to increased latency and energy consumption. Compared to previous methods, it can be observed our approach demonstrates greater efficiency in terms of latency and energy as K and C are scaled up. 

\subsubsection{Power Evaluation} 

\begin{table}[h]
\centering
\renewcommand{\arraystretch}{1.5}
\caption{Hardware cost and feature comparison. }
\resizebox{0.46\textwidth}{!}{
\begin{tabular}{c|cccc}
\hline
& MZI-ONN\cite{shen2017deep}           & Crosslight(\textit{DAC})\cite{9586161}            & Holylight(\textit{DATE})\cite{8715195}         & Ours       \\ \hline \hline
Programming times                 & \small{$\beta_a M N K $}                  &\small{$\operatorname{min}(N, B) K C$}                  &\small{$\operatorname{min}(N, B) K C$}                   &\small{$\operatorname{min}(N, B)$}            \\ \hline

Latency              & \small{$\beta_a$}                 & \small{$\left\lceil\frac{N C}{B\beta_t}\right\rceil$}                 &\small{$\left\lceil\frac{NC}{B}\right\rceil$}                  &\small{$\left\lceil\frac{N}{BK}\right\rceil$}              \\ \hline
Power                & \small{$\beta_p M N K$}                  &\small{$\frac{\operatorname{min}(N, B) K}{\beta_t}$}                   &\small{$\operatorname{min}(N, B) K$}                    &\small{$\operatorname{min}(N, B)$}           \\ \hline
Control   Complexity & High              & High              & High              & Low        \\ \hline
\end{tabular}
}
\label{table2}
\end{table}
\begin{table}[h]
\renewcommand{\arraystretch}{1.5}
\centering
\caption{Energy and delay performance of our proposed method. It is evaluated using 8 matrices with dimensions of 256 $\times$ 256 with only one basic matrix( reuse one matrix 8 times)}
\small
\resizebox{0.45\textwidth}{!}{%
    \begin{tabular}{c|c c|c c|c c}
    \hline\hline
     \multirow{2}{*}{Tile$=$N}&  \multicolumn{2}{c|}{64 }& \multicolumn{2}{c|}{256} & \multicolumn{2}{c}{1024} \\
     & Delay(ns) & Energy($\mu J$) & Delay(ns) & Energy($\mu J$) & Delay(ns) & Energy($\mu J$) \\
    \hline
    No Reuse &  217190 & 35.70 & 54297 & 9.68 & 13574 & 3.17 \\
    \hline
    Reuse  & 77490 & 12.50 & 20197 & 3.35 & 5874 & 1.06 \\
    \hline\hline
    \end{tabular}
}
\label{table3}
\end{table}

\begin{table}[t]
\definecolor{mygray}{gray}{.9}
\caption{Performance of the R\&B architecture.(W8A8 Quantization)\\ }
\vspace{-4mm}
\centering
\newcommand{\tabincell}[2]{\begin{tabular}{@{}#1@{}}#2\end{tabular}}
\resizebox{0.45\textwidth}{!}{
\begin{tabular}{cc>{\columncolor{mygray}}ccccc}
\toprule 
\multirow{3}{*}{Models} & \multirow{3}{*}{Datasets} &\multicolumn{2}{c}{\# Arch.}  &\multirow{3}{*}{\#P} & \multirow{3}{*}{\#E($\mu J$)} & \multirow{3}{*}{Acc(\%)}  \\ 
&& \cellcolor{white}\#Reuse Granularity & \tabincell{c}{\#Basic Matrix $\times$ \\ Reuse Times}  \\
\cmidrule{3-4}
\midrule
\multirow{10}{*}{MLP} & \multirow{5}{*}{MNIST} & \multicolumn{2}{c}{Baseline} & $0.36 \mathrm{~M}$ & 6.59 & 98.97 \\
                      &                        & \multicolumn{2}{c}{HolyLight(\textit{DATE})}  & $0.36 \mathrm{~M}$ & 5.92 & 98.81 \\
                      &                        & \multicolumn{2}{c}{CrossLight(\textit{DAC})} &$0.36 \mathrm{~M}$ & 5.27 & 92.93 \\
\cmidrule{3-7} 
                      &                        & \cellcolor{mygray} {layer-wise}&\cellcolor{mygray}$1\times6$ &\cellcolor{mygray} 0.16M &\cellcolor{mygray} \textbf{4.67} &\cellcolor{mygray} 98.33 \\
\cmidrule{2-7} 
                      & \multirow{5}{*}{FMNIST} & \multicolumn{2}{c}{Baseline} & $0.36 \mathrm{~M}$ & 6.59 & 89.51 \\
                      &                         & \multicolumn{2}{c}{HolyLight} & $0.36 \mathrm{~M}$ & 5.54 & 87.75 \\
                      &                         & \multicolumn{2}{c}{CrossLight} & $0.36 \mathrm{~M}$ & 4.72 & 89.95 \\
\cmidrule{3-7} 
                      &                         &\cellcolor{mygray} layer-wise &\cellcolor{mygray} $2\times3$ &\cellcolor{mygray} $0.23 \mathrm{~M}$ &\cellcolor{mygray} \textbf{4.67} &\cellcolor{mygray} 88.87 \\
\midrule

\multirow{5}{*}{VGG-13} & \multirow{5}{*}{CIFAR10} & \multicolumn{2}{c}{Baseline} & $9.42 \mathrm{~M}$ & 160.17 & 91.02 \\
                      &                        & \multicolumn{2}{c}{HolyLight} & $9.42 \mathrm{~M}$ & 128.15 & 91.27 \\
                      &                        & \multicolumn{2}{c}{CrossLight} & $9.42 \mathrm{~M}$ & 112.14 & 80.23 \\
\cmidrule{3-7} 
                      &                        &&\cellcolor{mygray} $4\times2$ &\cellcolor{mygray} $4.95 \mathrm{~M}$ &\cellcolor{mygray} \textbf{54.48} &\cellcolor{mygray}90.12 \\
                      &                       &\multirow{-2}{*}{layer-wise} &\cellcolor{mygray} $4\times4$ &\cellcolor{mygray} $4.95 \mathrm{~M}$ &\cellcolor{mygray} \textbf{100.95} &\cellcolor{mygray}90.65\\
\midrule

\multirow{5}{*}{ResNet-18} & \multirow{5}{*}{CIFAR10} & \multicolumn{2}{c}{Baseline} & $9.22 \mathrm{~M}$ & 184.27 & 93.33 \\
                      &                        & \multicolumn{2}{c}{HolyLight} &  $9.22 \mathrm{~M}$ & 128.02 & 93.16  \\
                      &                        & \multicolumn{2}{c}{CrossLight} &  $9.22 \mathrm{~M}$ & 99.67 & 84.75 \\
\cmidrule{3-7} 
                      &                         &&\cellcolor{mygray} $2\times4$ &\cellcolor{mygray} $3.10 \mathrm{~M}$ &\cellcolor{mygray} \textbf{75.61} &\cellcolor{mygray} 92.52 \\
                      &                        &\multirow{-2}{*}{layer-wise} &\cellcolor{mygray} $2\times8$ &\cellcolor{mygray} $3.10 \mathrm{~M}$ &\cellcolor{mygray} \textbf{136.42} &\cellcolor{mygray} 92.87 \\
\midrule
                      
\multirow{8}{*}{MLP-Mixer} & \multirow{9}{*}{CIFAR10} & \multicolumn{2}{c}{Baseline} & $0.68 \mathrm{~M}$ & 12.67 & 91.37 \\
                      &                        & \multicolumn{2}{c}{HolyLight} & $0.68 \mathrm{~M}$ & 10.93 & -- \\
                      &                        & \multicolumn{2}{c}{CrossLight} & $0.68 \mathrm{~M}$ & 9.37 & -- \\
\cmidrule{3-7} 
                      &                       && \cellcolor{mygray} $1\times8$ &\cellcolor{mygray} $0.15 \mathrm{~M}$ &\cellcolor{mygray} \textbf{7.51} &\cellcolor{mygray} 89.90 \\
                      &&&                         \cellcolor{mygray}$1\times12$ &\cellcolor{mygray} $0.15 \mathrm{~M}$ &\cellcolor{mygray} \textbf{11.04} &\cellcolor{mygray} 90.41\\
                      &&&                         \cellcolor{mygray}$2\times2$ &\cellcolor{mygray} $0.29 \mathrm{~M}$ &\cellcolor{mygray} \textbf{4.00} &\cellcolor{mygray} 89.56\\
                      &&&                         \cellcolor{mygray}$2\times3$ &\cellcolor{mygray} $0.29 \mathrm{~M}$ &\cellcolor{mygray} \textbf{5.65} &\cellcolor{mygray} 90.52\\
                      &&\multirow{-5}{*}{block-wise}&                         \cellcolor{mygray}$2\times4$ &\cellcolor{mygray} $0.29 \mathrm{~M}$ &\cellcolor{mygray} \textbf{7.29} &\cellcolor{mygray} 90.73\\

\bottomrule 
\end{tabular}
}
\noindent\footnotesize{\raggedright *baseline~refers~to~the~model~without~weight~sharing~or~other~techniques.}
\vspace{-4mm}
\label{table4}
\end{table}

Tuning each MRR to their desired value traditionally consumes a considerable amount of energy, particularly in large scale ONN applications. We evaluated our R\&B method on a silicon photonic accelerator with the properties shown in Table~\ref{table1}. The previous MRR ONN would use about 69\% more energy for computation compared to our R\&B method.

To provide a comprehensive evaluation of our method's performance, we tested it across a variety of datasets and ONN configurations. The results, summarized in Table~\ref{table3}, clearly demonstrate that the R\&B method not only maintains a level of accuracy comparable to conventional systems but also significantly enhances system efficiency. In terms of latency, our approach achieved an improvement of up to 57\%, with the settings demonstrated in Table~\ref{table3}. Moreover, it reduced energy consumption by at least 69\% using a 2 $\times$ 2 block-wise reused MLP-mixer, while maintaining comparable accuracy. The detailed energy consumption breakdown is shown in Fig~\ref{fig1}.

\vspace{{1mm}}
\subsection{Performance Evaluation}
\subsubsection{Effectiveness of the PRM}

To comprehensively investigate the effectiveness of PRM and OBU in enhancing the efficiency and performance of neural networks, we visualize the total model size and accuracy of the model with different reuse granularity and reuse times based on the minimal reusable component of the architecture in Table~\ref{table4}. In the table, the reuse times refer to the number of times a block or layer is reused within the model. For instance, in the MLP-Mixer model, a basic reusable component consists of a block that includes several layers focused on channel-mixing and token-mixing. Thus, We specifically implemented block-wise reuse in the MLP-Mixer to evaluate its impact on the model's footprint and performance.

The R\&B architecture significantly reduces the number of parameters and thus the programming frequency of MRRs. Across all experiments, PRM could reduce the model size by at least 34\% while maintaining a competitive level of accuracy. Specifically, our method shrinks the MLP-Mixer model by 42\% while keeping accuracy degradation within 0.64\% of the original. By reusing layers or blocks multiple times in combination with OBU, the weights can learn more varied and intricate patterns, thereby maintaining the model's accuracy. The R\&B architecture also significantly lowers energy consumption. For example, in the VGG-13 model on the CIFAR10 dataset, energy consumption decreases from 160.17$\mu J$ (Baseline) to 54.48$\mu J$ (layer-wise, 4×2 reuse). 

Overall, the integration of the PRM and OBU in the R\&B architecture maintained the model's accuracy close to its original level while significantly reducing energy consumption and delays associated with frequent recalibrations and the need for multiple MRR writings. This reduction is particularly important in scenarios where energy efficiency and quick response times are crucial.

\subsubsection{Area Evaluation} The weight-sharing architecture substantially decreases the required number of MRR arrays, as demonstrated in Table~\ref{table4}. For instance, just one hardware block is necessary to implement MLP-Mixer on silicon photonics hardware with weight sharing, versus 8 hardware blocks without weight sharing, and only a 1.47\% drop in accuracy. Hence, sharing weights over blocks or layers greatly reduces model parameters, enabling silicon photonics accelerators to implement more complex neural network models with intricate connections and numerous parameters.
 
\subsubsection{Effect of Write Variation on Energy Consumption and Aging Degradation}
Tuning and calibrating MRRs consume considerable energy. Specifically, the trimming power for remedy 1 $nm$ shift in the MRR's resonance wavelength reaches 240 $mW$ of thermal tuning power \cite{8382285}. In large-scale ONNs, this culminates in a prohibitively high total energy consumption on the order of tens of watts. The weight-sharing architecture helps to reduce up to 69\% energy consumption by reducing the reprogramming and calibration frequency.

Besides, aging variations present significant challenges. Long-term operation with frequent temperature changes will cause MRR degradation, which leads to resonance wavelength drifts, and a decrease in the Q-factor \cite{8752130}. As shown in Table~\ref{table2}, Our system requires programming times and power consumption of only $\operatorname{min}(N,B)$, whereas other systems need additional parameters K and C. Therefore, our method gains more hardware efficiency advantage with low latency and low control complexity compared to previous MRR-ONN. However, implementing advanced ANN models that require both large K and C on silicon photonic platforms inevitably necessitates multiple heat-tuning operations. This results in significant fluctuations in elevated voltage biases or temperatures, exacerbating the aging problem of MRRs. With at least 22\% fewer parameters and a 40\% energy saving (1 $\times$ 8 block-wise MLP-mixer), our proposed technique reduces the need for tuning cycles, substantially mitigating aging-induced degradation and thereby enhancing the endurance of ONNs for practical machine-learning applications.

\subsubsection{Ablation Study} 
We carried out an evaluation of our method using the CIFAR10 dataset. For this, we employed MLP-Mixer models with a range of configurations, as detailed in Table~\ref{table5}. When only OBU is applied to the model without using PRM, a single shuffling operation shows an improvement in accuracy, increasing by 0.41\%. The failure of a single transpose operation might be due to the unique inner transpose operation in the MLP-mixer. Compared to the basic model using PRM, applying the shuffle or transpose technique improves accuracy by up to 3.16\% and 1.77\%, respectively. The results we obtained underscore the efficacy of our OBU design. These units facilitate optical transposing and electrical shuffling and significantly enhance model performance without adding to the overhead.  Notably, our blend units can be efficiently incorporated into the existing MRR-ONN architectures with negligible overhead. Our experiments with the CIFAR10 dataset illustrate the capability of this method to amplify model performance for computer vision tasks while preserving an efficient and compact model architecture.



\begin{table}[h]
\renewcommand{\arraystretch}{0.9}
\definecolor{mygray}{gray}{.9}
\caption{ Ablation Study on CIFAR10 dataset. The row in gray is the baseline.}
\vspace{-2mm}
\centering
\begin{adjustbox}{width=0.45\textwidth}
    \begin{tabular}{cc|ccc|c}
    \hline \hline
     \multirow{2}{*}{Dataset} &\multirow{2}{*}{Model} &\multicolumn{3}{c|}{Method}  &\multirow{2}{*}{ACC(\%)}   \\ \cline{3-5} 
    && \multicolumn{1}{c}{reuse} &shuffle & \multicolumn{1}{c|}{transpose} \\ 
    \hline
    \multirow{7}{*}{CIFAR10} &\cellcolor{mygray}\begin{tabular}[c]{@{}c@{}}MLP-Mixer\end{tabular}  &\cellcolor{mygray}\XSolidBrush&\cellcolor{mygray}\XSolidBrush&\cellcolor{mygray}\XSolidBrush& \cellcolor{mygray} 91.37 \\ 
    \cline{2-6}
    & \multirow{6}{*}{\begin{tabular}[c]{@{}c@{}}MLP-Mixer \\ (2 $\times$ 4)\end{tabular}} 
    &\CheckmarkBold &  & & 87.67 \\
    &&\CheckmarkBold & \CheckmarkBold & & 90.83 \\ 
    &&\CheckmarkBold &  &\CheckmarkBold & 89.44 \\ 
    &&& \CheckmarkBold &  & 91.78 \\ 
    &&&  & \CheckmarkBold & 90.91 \\ 
    &&& \CheckmarkBold & \CheckmarkBold & 87.81 \\ 
    \hline\hline
    \end{tabular}
\end{adjustbox}
\label{table5}
\end{table}

\vspace{-2mm}
\section*{CONCLUSION}

In this research, we propose a novel approach to optical computing design that outperforms existing designs in terms of both power consumption and inference delays.  We devise PRM and OBU to allow efficient block-wise or layer-wise weight reusing. Specifically, our R\&B architecture significantly diminishes the necessity for frequent reprogramming and intricate calibration processes, thereby streamlining the integration of sophisticated ANN architectures into silicon photonics platforms. Our experimental results demonstrate that this architecture not only aligns well with the theoretical expectations but also presents a viable and effective framework for deploying machine learning algorithms in practical settings using silicon-photonics technology. Our R\&B architecture, with its low latency and low power consumption, is a highly promising candidate for future machine learning and artificial intelligence applications, leveraging the advanced capabilities of silicon photonics.

\vspace{-2mm}
\section*{Acknowledgement}
This work was partially sponsored by the Zhejiang Lab Open Research Project (NO.117008-AB2201). This work was also partially supported by the Guangzhou Municipal Science and Technology Project (No.2023A03J0013).


\bibliographystyle{unsrt}
\bibliography{ref}

\end{document}